\documentclass[twocolumn,nofootinbib,showpacs,preprintnumbers,superscriptaddress] {revtex4}

\usepackage{amssymb,amsfonts,amsmath,graphicx}

\def\mathbi#1{\textbf{\em #1}}

\newcommand{\VEV}[1]{\langle #1 \rangle}
\newcommand{\mpl}{m_{\rm Pl}}
\newcommand{\calO}{{\cal O}}

\newcommand\krh{k_{\rm reh}}
\newcommand\kkd{k_{\rm kd}}
\newcommand\kfr{k_{\rm fr}}

\newcommand{\Trh}{T_{\rm reh}}
\newcommand\Tkd{T_{\rm kd}}

\begin{document}

\title{WIMP isocurvature perturbation and small scale structure}

\author{Ki-Young Choi}
\email{kiyoungchoi@kasi.re.kr}
 \affiliation{Korea Astronomy and Space Science Institute, Daejeon 305-348, Korea}

\author{Jinn-Ouk Gong}
\email{jinn-ouk.gong@apctp.org}
\affiliation{Asia Pacific Center for Theoretical Physics, Pohang, 790-784, Korea}
\affiliation{Department of Physics, Postech, Pohang 790-784, Korea}

\author{Chang Sub Shin}
\email{changsub@physics.rutgers.edu}
\affiliation{Department of Physics and Astronomy, Rutgers University, Piscataway NJ 08854, USA}

\begin{abstract}

The adiabatic perturbation of dark matter is damped 
during the kinetic decoupling due to the collision with 
relativistic component on sub-horizon scales. 
However the isocurvature part is free from damping and 
could be large enough to make a substantial contribution 
to the formation of small scale structure. 
We explicitly study the weakly interacting massive particles 
as dark matter with an early mater dominated period 
before radiation domination and show that 
the isocurvature perturbation is generated during the phase transition 
and leaves imprint in the observable signatures for small scale structure.

\end{abstract}


\preprint{APCTP-Pre2015-019, RUNHETC-2015-06}

\maketitle

{\it Introduction}.\quad
The formation of large scale structure is consistent with 
non-relativistic dark matter (DM) independent of its nature. 
Small scale structure, however, depends on the microphysics 
of DM and the corresponding evolution in the early 
universe~\cite{Erickcek:2011us,Barenboim:2013gya,Fan:2014zua,Loeb:2005pm}. 
For weakly interacting massive particles (WIMPs), 
the kinetic decoupling is a crucial stage to determine 
the size of smallest object~\cite{Hofmann:2001bi,Green:2005fa}: 
during the process of kinetic decoupling collisional damping smears out 
the inhomogeneities below the corresponding damping scale. 
After kinetic decoupling WIMPs can move freely and this leads to 
additional damping below the free streaming scale. 
For neutralino DM, the kinetic decoupling scale is set 
when the temperature is 10 MeV - 1 GeV
for the mass between 100 GeV and TeV~\cite{Bringmann:2009vf}.

In radiation dominated era (RD), while the ``adiabatic'' 
component of DM perturbation on sub-horizon scales 
experiences oscillations followed by collisional 
damping~\cite{Ma:1995ey}, the isocurvature perturbation 
between DM and radiation,
\begin{equation}
S \equiv 3H \left( \frac{\delta\rho_m}{\dot\rho_m} - 
\frac{\delta\rho_r}{\dot\rho_r} \right) 
= \delta_m - \frac{3}{4}\delta_r \, ,
\end{equation}
remains constant without damped oscillations~\cite{Peebles87,Hu:1994jd}. 
This property was used to explain large scale structure with 
baryon isocurvature perturbation~\cite{Peebles87}, 
which is ruled out now by the adiabatic constraint from 
the cosmic microwave background (CMB)~\cite{Ade:2015xua}. 
However, large isocurvature perturbation on small scales is 
not constrained by the CMB observations and can give 
observable signatures in small scale structure.

In this article, we show how large isocurvature perturbation of 
WIMPs can be generated for scales 
that enter the horizon before the kinetic decoupling. 
If $S=0$ at the onset of RD, it remains so during kinetic equilibrium.
Instead, if an early matter dominated era precedes RD, 
a sizable amount of $S$ can be generated.
We note that this isocurvature perturbation will not be damped 
even if the kinetic decoupling happens after the transition to RD.

{\it Dark matter in non-thermal background}.\quad
In the early universe, it happens often that the energy density 
of the universe is dominated by a non-relativistic matter 
which subsequently decays into relativistic particles. 
This non-relativistic matter includes a coherently oscillating 
scalar field like an inflaton, or massive fields which decay 
very late, such as curvaton, moduli and so on. 
As an illustration, we consider this dominating non-relativistic 
matter as a scalar $\phi$ with a decay rate $\Gamma_\phi$. 
Accordingly, we call the epoch during which $\phi$ dominates 
the energy density as the scalar dominated era (SD). 
In the background, then there are three species of fluid: 
$\phi$, radiation and DM. 
Their evolutions are governed by the continuity equations,
\begin{align}
\label{eq:rho_phi}
\dot\rho_\phi + 3 H \rho_\phi & = - \Gamma_\phi \rho_\phi \, , 
\\
\label{eq:rho_r}
\dot\rho_r + 4 H \rho_r & = (1-f_m) \Gamma_\phi \rho_\phi 
+ \frac{\VEV{\sigma_a v}}{M} \left[ \rho_m^2 - \left( \rho_m^{\rm eq} \right)^2 \right] \, ,
\\
\label{eq:rho_m}
\dot\rho_m+3H \rho_m & =  f_m\Gamma_\phi \rho_\phi 
- \frac{\VEV{\sigma_a v}}{M} \left[ \rho_m^2 - \left( \rho_m^{\rm eq} \right)^2 \right] \, ,
\end{align}
where $M$ is the mass of the DM particle, 
$f_m$ is the fraction of the decay of $\phi$ into DM, 
$\VEV{\sigma_a v}$ is the thermal averaged annihilation cross section of DM 
and $\rho_m^{\rm eq} \approx M^4(2\pi M/T)^{-3/2} \exp(-M/T)$ 
is the energy density of DM in thermal equilibrium. 
Here radiation is the relativistic particles thermalized quickly 
when produced from the decay of $\phi$, and thus 
the temperature $T$ is properly defined by its energy density 
$\rho_r = \pi^2g_*T^4/30$ with $g_*$ being the effective degrees 
of freedom of the relativistic particles in thermal equilibrium. 
The reheating temperature is then approximately given by 
$\Trh \approx (\pi^2g_*/90)^{-1/4}\sqrt{\mpl\Gamma_\phi}$. 
For successful big bang nucleosynthesis, we require that 
$\Trh$ must be larger than $\calO$(MeV)~\cite{Hannestad:2004px}.

\begin{figure}[!t]
\begin{center}
\begin{tabular}{c} 
 \includegraphics[width=0.45\textwidth]{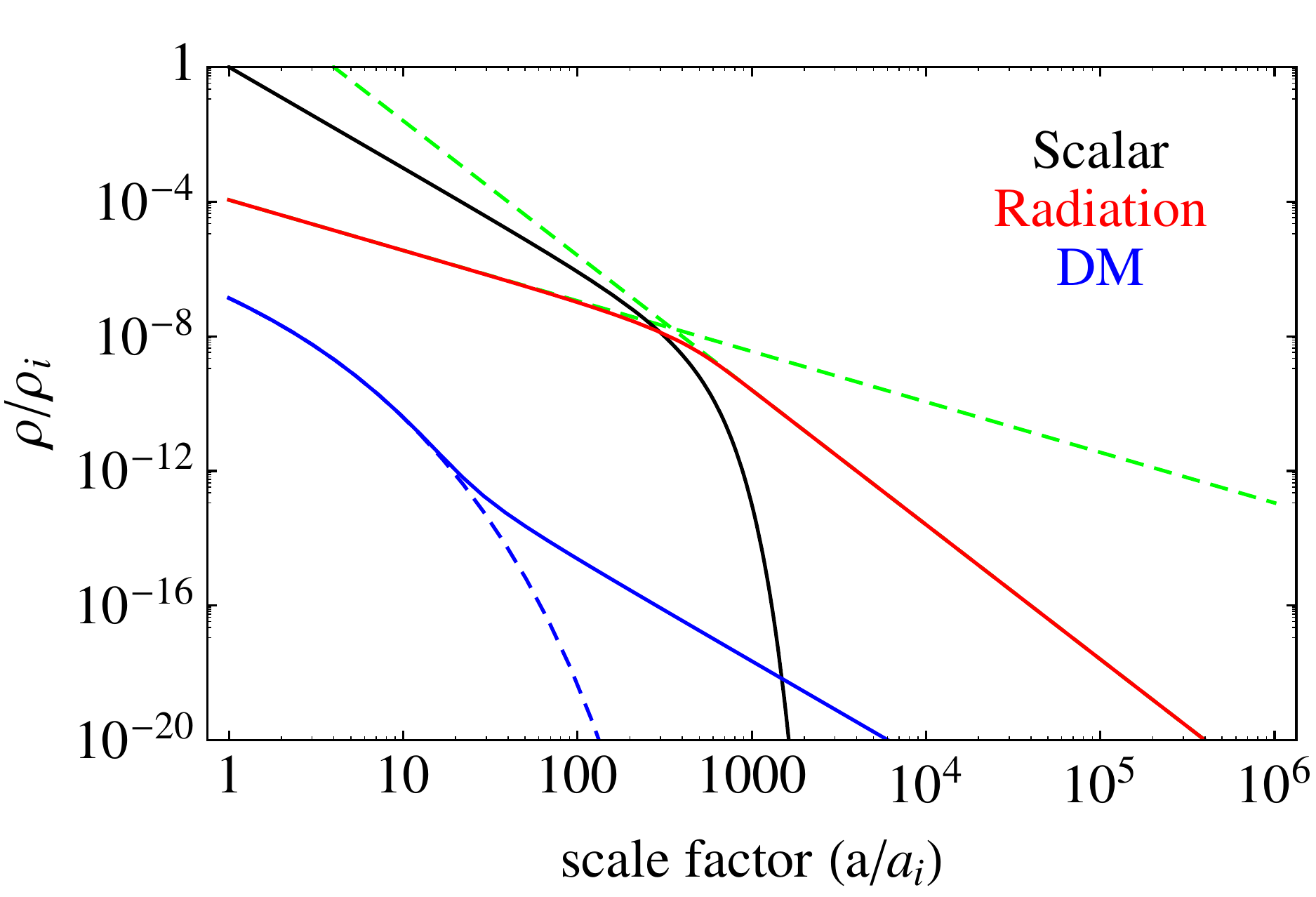}
 \end{tabular}
\end{center}
\caption{
The evolution of the energy densities of the scalar (black), radiation (red) 
and DM (blue) respectively with respect to the initial total energy density. 
Blue dashed line is the equilibrium energy density of WIMP, 
and green dashed lines denote the asymptotic behavior of radiation energy density.
DM freezes out at $a/a_i\simeq 20$ and RD starts from $a/a_i\simeq 300$.
} 
\label{fig:rhos}
\end{figure}

While radiation is produced directly from the decay of $\phi$, 
DM can be produced in several different ways~\cite{Baer:2014eja}. 
For simplicity, we assume that DM is produced only from radiation 
by scatterings and set $f_m=0$. Even in this case, 
a sizable amount of DM can be produced from thermal plasma. 
If the interaction of DM with plasma is large enough, 
they could be in thermal equilibrium. WIMP is one such example, 
which is intimately coupled to the relativistic plasma and decoupled 
when $T/M \sim 1/20$, depending on the annihilation cross section 
$\VEV{\sigma_a v}$~\cite{Lee:1977ua}. 
The freeze-out may happen during SD or RD after the scalar decay. 
For the latter case, there will be no difference from the thermal WIMP 
in the standard scenario. Therefore, in our study, we will focus on 
the case that WIMPs are decoupled during SD.

In Figure~\ref{fig:rhos}, we show the evolution of the background 
energy densities of $\phi$, radiation and DM by solving  
\eqref{eq:rho_phi}-\eqref{eq:rho_m}. During SD, $\rho_r$ scales as
$\rho_r\propto a^{-3/2}$ due to the continuous production 
from the scalar decay and thus the effective equation of state 
during SD is $-1/2$.
DM is frozen during SD, and its energy density decreases 
simply proportional to $a^{-3}$ after then.
However the interactions by collisions continue until RD.

\begin{figure*}[!t]
\begin{center}
 \begin{tabular}{ccc} 
  \includegraphics[width=0.31\textwidth]{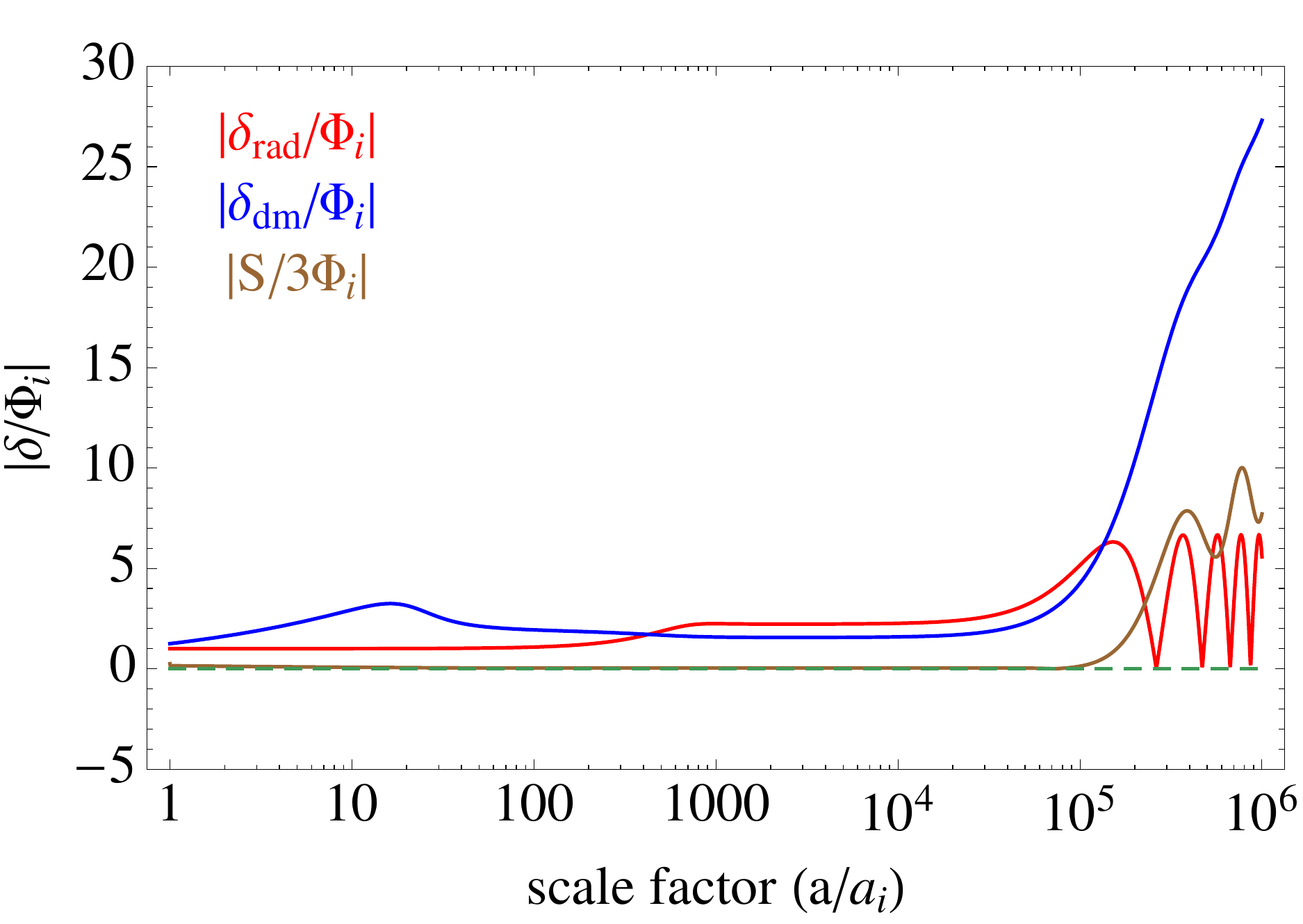}
  &
  \includegraphics[width=0.31\textwidth]{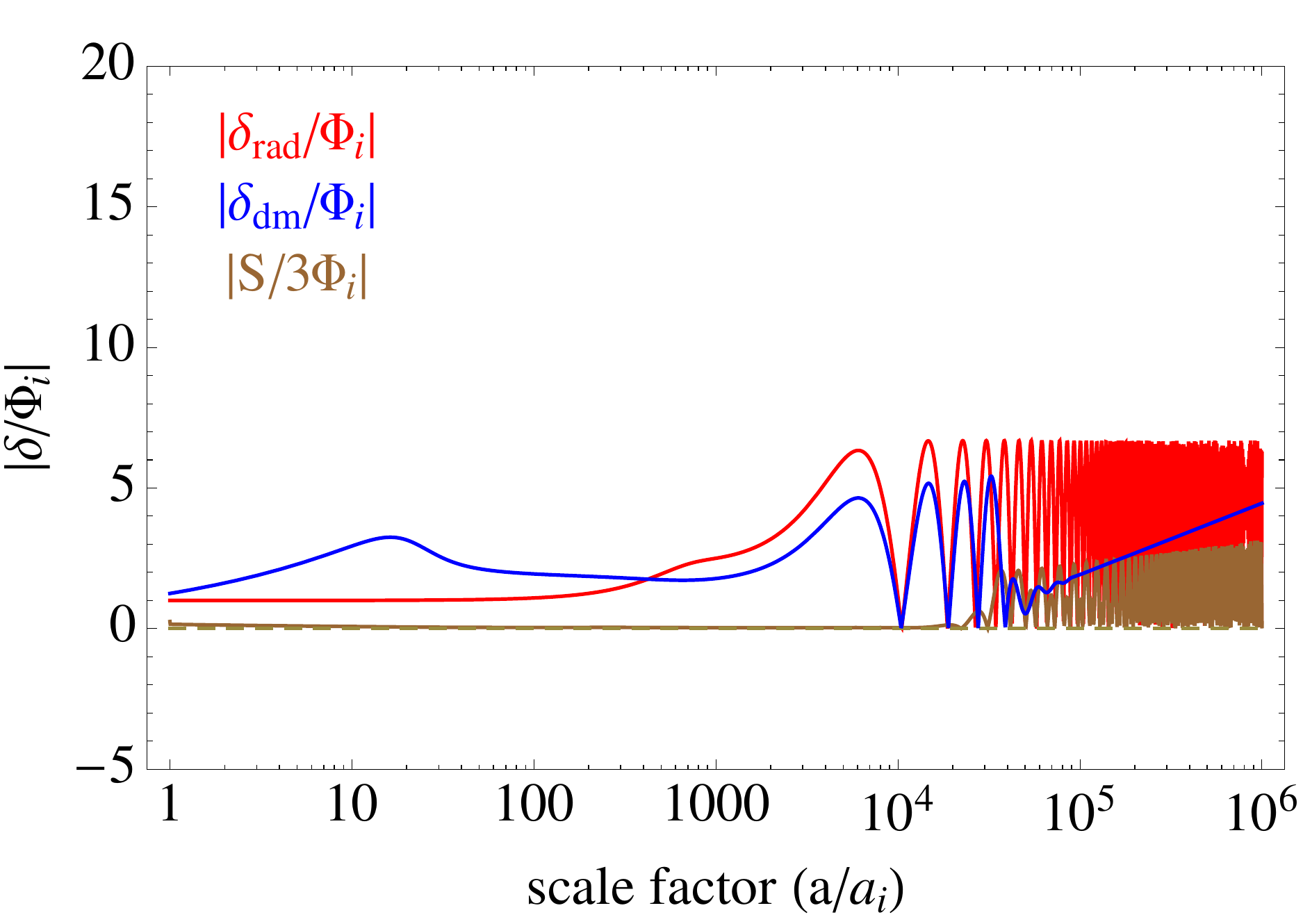}
  &
  \includegraphics[width=0.31\textwidth]{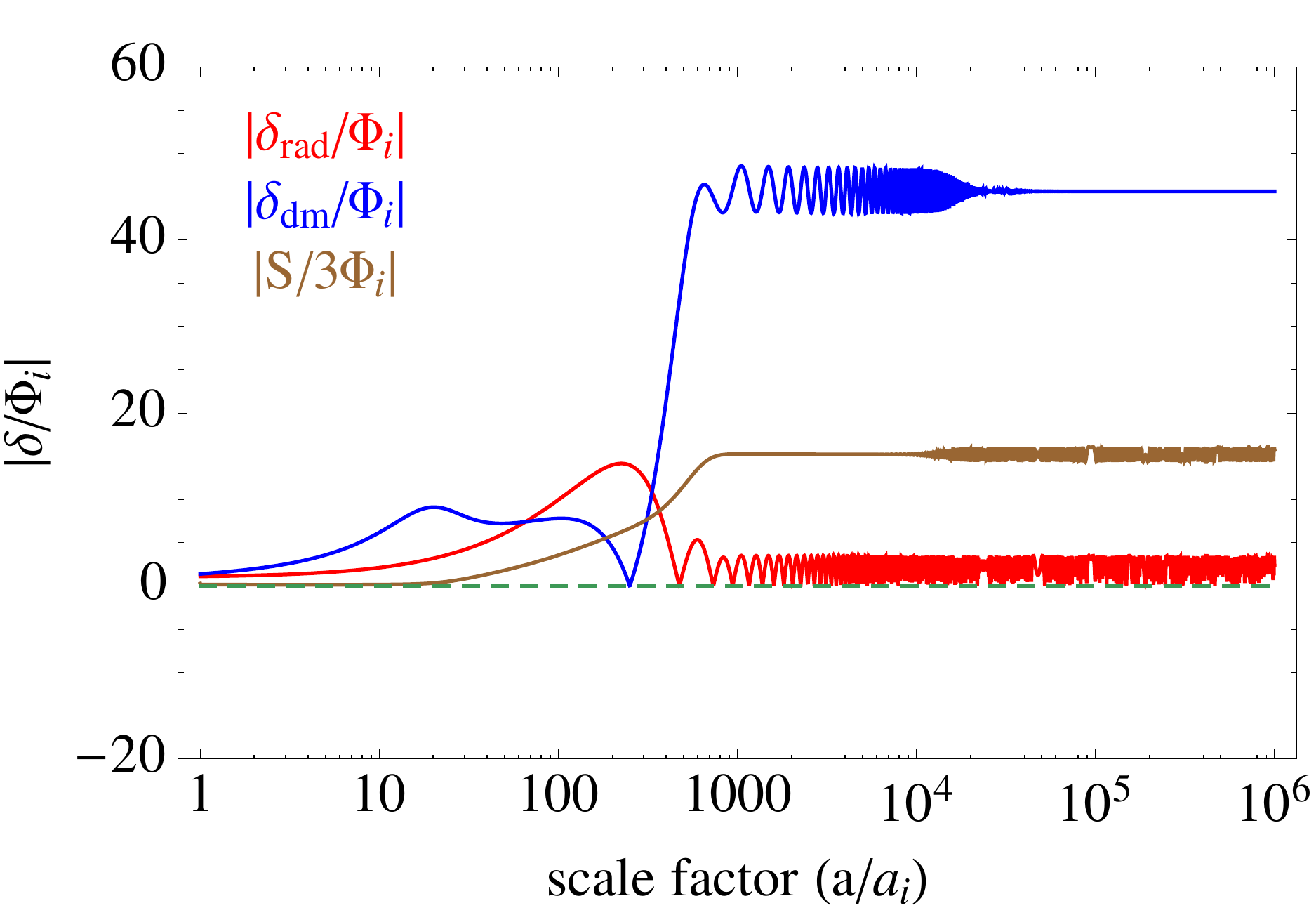}
 \end{tabular}
\end{center}
\caption{ 
The evolution of the density contrast of the radiation (red), 
DM (blue) and the isocurvature perturbation (brown) 
with respect to the initial gravitational potential 
for $ \kkd^{-1}<k^{-1} $ ($k = 0.1\,\kkd$, left), 
$ \krh^{-1}< k^{-1} < \kkd^{-1}$ ($k = 5\,\kkd = 0.5\, \krh$, middle) 
and $\kfr^{-1} < k^{-1} <\krh^{-1}$ ($k =50\, \kkd = 5\, \krh = 0.8\kfr$, right). 
We have set $M= 5 \, {\rm TeV}$, $\Trh=0.1\, {\rm GeV}$ and $\Tkd=0.01\,{\rm GeV}$.
} 
\label{fig:delta}
\end{figure*}

{\it Evolution of perturbations}.\quad
Now we consider the evolution of perturbations.
For this, we use the Newtonian gauge with the metric
\begin{equation}\label{metric}
ds^2 = -(1+2\Phi)dt^2 + a^2 (1-2\Psi)\delta_{ij} dx^idx^j \, .
\end{equation}
The perturbation equations can be derived from 
the Boltzmann equation for each component 
($\alpha=\phi$, $r$ and $m$) and they are given by~\cite{Ma:1995ey,Dodelson}
\begin{widetext}
\begin{align}
\label{eq:delta}
\dot\delta_\alpha+ (1+w_\alpha)\frac{\theta_\alpha}{a} - 3(1+w_\alpha)\dot\Psi & = 
\frac{1}{\rho_\alpha} \left( \delta Q_\alpha - Q_\alpha\delta_\alpha +Q_\alpha\Phi \right) \, ,  
\\ 
\label{eq:theta}
\dot\theta_\alpha+ (1-3w_\alpha)H\theta_\alpha + \frac{\Delta\Phi}{a} + \frac{w_\alpha}{1+w_\alpha}\frac{\Delta\delta_\alpha}{a} & = \frac{1}{\rho_\alpha} \left[ \frac{\partial_iQ_{(\alpha)}^i} {1+w_\alpha}-Q_\alpha\theta_\alpha\right] \, ,
\end{align}
\end{widetext}
where $\theta_\alpha \equiv \nabla\cdot\mathbi{v}_\alpha = \partial_i v_\alpha^i$ 
is the velocity divergence field, $w_\phi=w_m=0$ and $w_r=1/3$. 
At leading order of $T/M$, the energy-momentum transfer functions 
$Q_\alpha$ and $\delta Q_\alpha$ 
can be calculated from the Boltzmann equation as
\begin{align}
Q_\phi &= -\Gamma_\phi \rho_\phi \, , 
\\
Q_r &= \Gamma_\phi \rho_\phi + \frac{\langle \sigma_a v\rangle}{M} 
\left[ \rho_m^2 - (\rho_m^{\rm eq})^2 \right] \, ,
\\ 
Q_m & = - \frac{\langle \sigma_a v\rangle}{M} 
\left[ \rho_m^2 - (\rho_m^{\rm eq})^2 \right] \, ,
\\
\delta Q_\phi &=-\Gamma_\phi\rho_\phi \delta_\phi,
\\
\delta Q_r &= \Gamma_\phi \rho_\phi\delta_\phi + 
\frac{2\langle \sigma_a v\rangle}{M}
\left[ \rho_m^2\delta_m - (\rho_m^{\rm eq})^2
\frac{M}{T} \frac{\delta_r}{4} \right] \, , 
\\
\delta Q_m  & = -\frac{2\langle \sigma_a v\rangle}{M}
\left[ \rho_m^2\delta_m - (\rho_m^{\rm eq})^2 
\frac{M}{T} \frac{\delta_r}{4} \right] \, ,
\end{align}
and $\partial_iQ_{(\alpha)}^i$ by
\begin{align}
\partial_iQ_{(\phi)}^i & = - \Gamma_\phi \rho_\phi \theta_\phi  
\\ 
\partial_iQ_{(r)}^i & = \Gamma_\phi \rho_\phi \theta_\phi 
+ \frac{\langle \sigma_a v\rangle}{M}
\left[ \rho_m^2 \theta_m
- \frac{4}{3} (\rho_m^{\rm eq})^2 \frac{M}{2\pi T}^{1/2}\theta_r \right]  
\nonumber\\
& \quad
- c_e\frac{\langle\sigma_e v\rangle}{M}\rho_m \rho_r   
\left( \theta_r - \theta_m \right) \, ,
\\
\partial_iQ_{(m)}^i & =
-\frac{\langle \sigma_a v\rangle}{M}
\left[ \rho_m^2 \theta_m
- \frac{4}{3}(\rho_m^{\rm eq})^2
\frac{M}{2\pi T}^{1/2}\theta_r\right] 
\nonumber\\
& \quad + c_e\frac{\langle\sigma_e v\rangle}{M}  \rho_m \rho_r  
\left( \theta_r  - \theta_m \right) \, ,
\end{align}
where we have put $f_m=0$.
In the above equations, we have included the elastic scattering 
cross section between radiation and DM $\sigma_e$ 
which keeps DM and radiation in kinetic equilibrium 
until they decouple at $\Tkd$ set by 
$c_e\VEV{\sigma_ev}\rho_r/M|_{T=\Tkd} = H(\Tkd)$, 
with $c_e = \calO(1)$ being dependent on model 
due to the different momentum dependence of $\sigma_e$.

The 00 component of the perturbed Einstein equation 
governs the evolution of the metric perturbations,
\begin{equation}
\frac{\Delta}{a^2}\Psi - 3H \left( \dot\Psi + H\Phi \right) = 
\frac{1}{2\mpl^2} \left( \rho_\phi\delta_\phi + \rho_r\delta_r + \rho_m\delta_m \right) \, .
\end{equation}
In the absence of the anisotropic tensor, we can set 
$\Phi = \Psi$ which then closes the above set of equations.
This is possible since $\phi$ and radiation which dominate 
the energy density are isotropic in our setup.
Note that the effects of the anisotropic shear and 
non-vanishing sound speed of DM, $c_s \sim \sqrt{T/M}$, 
can be important after kinetic decoupling for scales 
smaller than the free streaming length $\kfr^{-1}$. 
In~\cite{Bertschinger:2006nq}, it is shown that 
when the free streaming length is much shorter than the scale 
$\kkd^{-1}$ that enter the horizon at the moment of kinetic decoupling, 
we can take an approximation that solving the Boltzmann equations first 
in perfect fluid limit while maintaining the elastic scattering, 
and then multiplying the solution by the Gaussian suppression term. 
Actually this limit is also physically interesting, 
because two different damping scales can be more clearly distinguished.

In this article, we consider the hierarchies among scales as 
$\kfr^{-1} <  \krh^{-1} < \kkd^{-1}$, where $\krh^{-1}$ is the 
scale that enters the horizon at $T=\Trh$. 
This means that the free streaming scale enters the horizon 
during SD and that kinetic decoupling occurs during RD.
The large hierarchy between $\kfr^{-1}$ and $\kkd^{-1}$ 
can be obtained when $M$ is big enough 
while the elastic scattering is mediated by a field much lighter than DM.
In this case, the freeze-out abundance also could be large, 
but the subsequent dilution by entropy injection from the scalar decay 
can provide the correct amount of the present  DM density~\cite{Choi:2008zq,Arcadi:2011ev}. 
For WIMP, we find~\cite{Kolb:1990vq}
\begin{align}
\kkd^{-1} & = 0.86 \frac{10 \, {\rm MeV}}{\Tkd} \left( \frac{g_{*s}}{10.75} \right)^{1/3} 
\left( \frac{10.75}{g_*} \right)^{1/2} {\rm pc} \, ,
\\
\krh^{-1} & = \kkd^{-1} \frac{\Tkd}{\Trh} \, , 
\\
\label{eq:kfr}
\kfr^{-1} & = \int_{t_{\rm kd}}^{t_0} \frac{dt}{a}c_s 
\approx \kkd^{-1} \sqrt{\frac{\Tkd}{M}} \log \left( \frac{\Tkd}{T_{\rm eq}} \right) \, ,
\end{align}
where $g_{*s}$ is the effective number of light species for entropy and 
$T_{\rm eq} = \calO({\rm eV})$ is the temperature at matter-radiation equality.

In Figure~\ref{fig:delta}, we show the evolution of perturbations 
on three different scales. During SD, the perturbations are adiabatic 
on super-horizon scales since both radiation and DM are 
produced from a single source $\phi$, 
which set the initial values of perturbations as 
$\delta_\phi(a_i) = 2\delta_r(a_i) = -2\Phi_i$ and 
$\delta_m(a_i) \approx M\delta_r(a_i)/(4T_i)$, 
with $T_i$ being determined from $\rho_r(a_i)$. 
During the transition from SD to RD, $\Phi$ rescales 
from $\Phi_i$ to $10\Phi_i/9$  on super-horizon scales and 
accordingly $\delta_r$ changes from $-\Phi_i$ to $-2(10/9)\Phi_i$. 
Meanwhile, at early times when DM is in thermal (chemical) equilibrium, 
$\delta_m \propto a^{3/8}$ and is reduced to $-5\Phi_i/3$ during RD 
which follows the adiabatic condition $\delta_m=3\delta_r/4$.

While for modes which enter the horizon after kinetic decoupling 
($\kkd^{-1}<k^{-1}$), $\delta_r$ oscillates and $\delta_m$ grows logarithmically 
as shown in the left panel of Figure~\ref{fig:delta}, 
for the modes which enter before kinetic decoupling 
($\krh^{-1} < k^{-1} < \kkd^{-1}$) $\delta_m$ oscillates together with $\delta_r$ 
and is damped, which is known as collisional damping. 
The non-vanishing sub-horizon entropy perturbation appears 
due to the damping of $\delta_m$ 
as shown in the middle panel of Figure~\ref{fig:delta}.

An interesting feature happens for the modes which enter the horizon  
during SD but after the free streaming scale enters  
($\kfr^{-1}< k^{-1} <\krh^{-1}$) as in the right panel of Figure~\ref{fig:delta}.
During the transition from SD to RD, 
$\delta_m$ does {\em not} follow $\delta_r$, 
and the isocurvature perturbation is generated. 
In this period, DM is no longer produced after chemical freeze-out 
and the number density is frozen while radiation is still being produced from $\phi$. 
The continuous entropy injection becomes 
the source of the isocurvature perturbation between DM and radiation. 
This perturbation still persists even after kinetic decoupling. 
Before calculating its analytic expression, 
we explicitly show why it is not damped, 
from the solution for $\delta_m$ during RD~\cite{Bertschinger:2006nq},
\begin{widetext}
\begin{align}
\label{eq:ddm_reh}
\delta_m & = \Delta_k \left[ 
\left( \frac{k}{\sqrt{3}aH} \right)^{-2} \cos \left( \frac{k}{\sqrt{3}aH} + \phi_k \right) 
- \left( \frac{k}{\sqrt{3}aH} \right)^{-3} \left( 1 - \frac{k^2}{3a^2H^2} \right) 
\sin \left( \frac{k}{\sqrt{3}aH} + \phi_k \right) \right.
\nonumber\\
&
\left. \hskip 1.3cm+ \int_{k/(\sqrt{3}aH)}^\infty \frac{\cos(x+ \phi_k)}{x} d x\right] 
 + A_k(t) \log \left( \frac{k}{\sqrt{3}aH} \right) + B_k(t) \, ,
\end{align}
\end{widetext}
where $\Delta_k$ and $\phi_k$ are $k$-dependent constants 
while $A_k(t)$ and $B_k(t)$ vary in time. 
Their time dependence is determined by the elastic scattering term as
\begin{equation}
\label{eq:friction}
\begin{split}
&\dot{A}_k + c_e\frac{\langle \sigma_ev\rangle \rho_r}{aM} A_k
 =  9 c_e\frac{\langle \sigma_ev\rangle \rho_r}{aM} 
\left[ \cos \left( \frac{k}{\sqrt{3}aH} + \phi_k \right) \right. 
\\
&\hskip 3.2cm\left.+ \frac{k}{2\sqrt{3} a H} 
\sin \left( \frac{k}{\sqrt{3} a H} + \phi_k \right)  \right] \, ,  
\\
&\dot{B}_k + \dot{A_k} \log \left( \frac{k}{\sqrt{3} aH} \right) = 0 \, .
\end{split}
\end{equation}
The values of $\Delta_k$, $\phi_k$, $A_k(t_{\rm reh})$ and $B_k(t_{\rm reh})$ are 
given at the onset of RD, and for adiabatic modes they are 
\begin{align}\label{eq:adiabatic}
& \Delta_k = -10\Phi_i \, , \quad \phi_k=0 \, ,\quad  A_k(t_{\rm reh}) = -10\Phi_i \, ,
\nonumber\\
&B_k^{\rm ad}(t_{\rm reh}) = -10\Phi_i \left (\gamma_E - \frac{1}{2} \right) \, ,
\end{align}
where $\gamma_E \approx 0.577$ is the Euler-Mascheroni constant. 
Then on super-horizon scales $k \ll aH$ we can recover $-5\Phi_i/3$ during RD.
For the modes which enters during RD ($\krh^{-1} < k^{-1} $),
the solution is~\cite{Bertschinger:2006nq} 
\begin{equation}
A_k \, , \, B_k^{\rm ad} \propto 
\exp \left[ -0.8 \left( \frac{k}{2\sqrt{3}\kkd} \right)^{\frac{4+n}{5+n}} \right] 
\end{equation}
for $\VEV{\sigma_ev} \propto T^{2+n}$, which clearly shows the damping 
for $k^{-1} \ll \kkd^{-1}$ due to the collision with radiation.

Here it is important to note that in \eqref{eq:friction} only $\dot{B}_k$ appears. 
The additional constant term to the adiabatic one is 
not damped away even in the kinetic equilibrium / decoupling periods.
As a result, for $k^{-1}\ll \kkd^{-1}$,  
$\delta_m$ is dominated by the isocurvature perturbation: 
$B_k = B_k^{\rm iso} + B_k^{\rm ad}\simeq B_k^{\rm iso}$.

{\it Generation of isocurvature perturbation}.\quad
For the modes that enter the horizon during SD 
after chemical decoupling of DM, $\delta_\phi$ grows linearly,
\begin{equation}
\label{eq:deltaphi_SD}
\delta_\phi(a) = -2\Phi_i - \frac{2}{3}\Phi_i \left[ \frac{k}{a_iH(a_i)} \right]^2 \frac{a}{a_i} \, ,
\end{equation}
and then logarithmically during RD.
Meanwhile, $\delta_r$ grows during SD, 
since radiation is continuously produced from the decay of $\phi$. 
However, after the transition from SD to RD, 
this enhancement is lost and $\delta_r$ oscillates 
with heavily suppressed amplitude~\cite{Erickcek:2011us}.

During kinetic equilibrium, DM is tightly coupled to radiation, 
so that $\theta_m\approx \theta_r$. 
Ignoring the effect of DM annihilation 
the relevant equations for $\delta_m$ and $\delta_r$ are, 
from \eqref{eq:delta},
\begin{align}
\dot\delta_m & \approx - \frac{\theta_r}{a } \, ,
\\
\label{eq:deltar-RD}
\dot\delta_r & \approx - \frac{4}{3}\frac{\theta_r}{a} 
+ \frac{\Gamma_\phi\rho_\phi}{\rho_r} \left(  \delta_\phi - \delta_r \right) \, ,
\end{align}
where we have neglected $\calO(1)$ contribution.
From SD to the transition period, both $\delta_r$ and $\Phi$ are 
sub-dominant compared to $\delta_\phi$, and
$\rho_r \approx 2\Gamma_\phi\rho_\phi/5H$. 
Then the isocurvature perturbation is
\begin{equation}
S(t_{\rm reh}) \approx   
-\frac{3}{4} \int_{t_i}^{t_{\rm reh}} dt \frac{\Gamma_\phi\rho_\phi \delta_\phi}{\rho_r}
\approx \frac{5}{4} \Phi_i \left( \frac{k}{\krh} \right)^2 \, .
\end{equation}
As can be read from \eqref{eq:deltar-RD}, unlike $\delta_m$, 
$\delta_r$ is sourced by both $\theta_r$ and $\delta_\phi$ 
because there is steady production of radiation from $\phi$. 
The corresponding isocurvature part becomes $ B_k^{\rm iso}$.

\begin{figure}[!t]
\begin{center}
 \begin{tabular}{cc} 
  \includegraphics[width=0.45\textwidth]{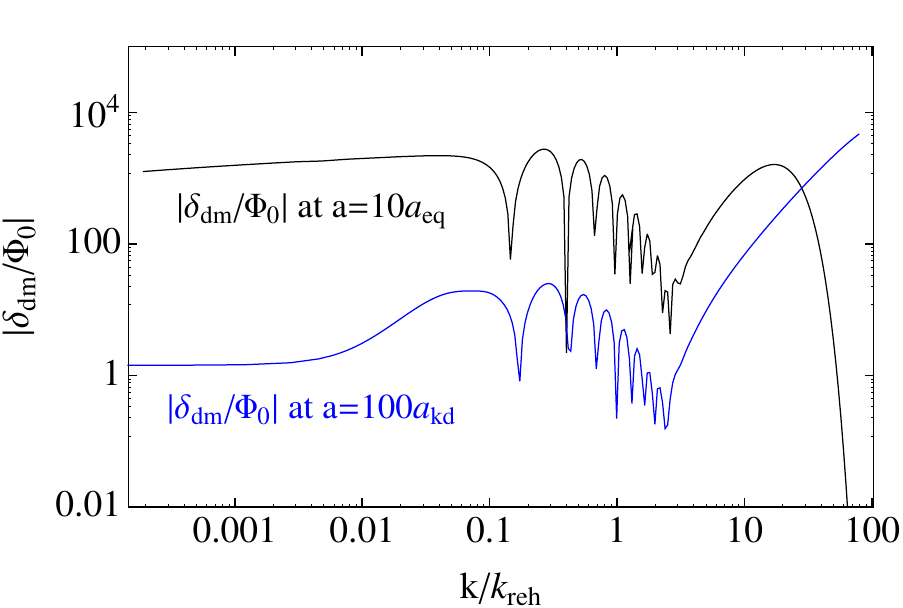}
 \end{tabular}
\end{center}
\caption{Density contrast of DM with $M= 5 \, {\rm TeV}$, 
$\Trh=0.1\, {\rm GeV}$ and $\Tkd=0.01\,{\rm GeV}$.
}
\label{fig:delta_dm}
\end{figure}

While the isocurvature perturbation can avoid the damping 
due to the collision, the diffusion by the free streaming still exist. 
Considering the damping effect due to free streaming, 
as discussed before we may add a Gaussian suppression factor 
to $\delta_m$ as
\begin{equation} 
\label{eq:deltasmall_freestream}
\delta_m \approx \exp \left( -\frac{k^2}{2\kfr^2} \right) 
\frac{5}{4} \Phi_i \left(\frac{k}{\krh}\right)^2 \, ,
\end{equation} 
where the free streaming scale $\kfr^{-1}$ is estimated as \eqref{eq:kfr}. 
Based on these results, it is straightforward to calculate 
the evolution of perturbation during the subsequent matter dominated era, 
the transfer function and the mass function. 
In Figure~\ref{fig:delta_dm}, we show $\delta_m$ at later stages at 
$a= 100\,a_{\rm kd}$ and $10\, a_{\rm eq}$.

{\it Implications}.\quad
At the heart of our finding is that despite of the damping of the conventional adiabatic 
perturbation of DM due to a large elastic scattering rate between DM and the standard 
model particles, the DM isocurvature perturbation survives the collisional damping 
until kinetic decoupling. This unsuppressed perturbation on small scale can give rise
to a large number of DM clumps, such as compact mini 
haloes~\cite{Diemand:2005vz,Bringmann:2011ut}.

Since DM can annihilate efficiently in the clumps, these haloes can serve as the sources of 
highly luminous gamma rays which can be well observed with the ongoing or future gamma-ray 
telescope like Fermi-LAT~\cite{Bertoni:2015mla} or Cerenkov Telescope Array~\cite{Carr:2015hta}.
They can be also the sources of neutrinos~\cite{Yang:2013dsa}, detectable by  
IceCube~\cite{Aartsen:2013dxa}. Furthermore they can leave an imprint in the CMB 
by changing the reionization history of the Universe~\cite{Berezinsky:2014wya}, produce a 
microlensing light curve~\cite{Ricotti:2009bs,Li:2012qha}, or change the direct detection 
rate~\cite{Kamionkowski:2008vw}.

The requisite for a large enough DM isocurvature perturbation is a sufficient hierarchy between 
DM freeze-out and reheating to have a long enough early MD. This is easily realized with a
low reheating temperature, which happens ubiquitously in many theoretical models when 
the heavy particle dominates and decays in the early Universe. Those models include 
the neutralino DM in the low reheating temperature~\cite{Gelmini:2006pw,Roszkowski:2014lga}  
and the scenario of decaying heavy particle such as moduli, gravitino~\cite{Kohri:2005ru}, or 
axino~\cite{Choi:2008zq}. Considering both astrophysical and cosmological observations and
DM theories should give more information about the early history of the universe before BBN
and the properties of DM.

{\it Acknowledgments}.
JG acknowledges the Max-Planck-Gesellschaft, the Korea Ministry of Education, Science and Technology, Gyeongsangbuk-Do and Pohang City for the support of the Independent Junior Research Group at the Asia Pacific Center for Theoretical Physics. JG is also supported by a Starting Grant through the Basic Science Research Program of the National Research Foundation of Korea (2013R1A1A1006701). 
CSS is supported in part by DOE grants DOE-SC0010008, 
DOE-ARRA-SC0003883, and DOE-DE-SC0007897.

\end{document}